\RequirePackage{fix-cm}
\documentclass[smallextended]{svjour3}       
\smartqed  
\usepackage{graphicx}
\usepackage{amsmath}
\usepackage{color}
\begin{document}

\title{ Some remarks on modularity density \thanks{Financial support by SUTD-MIT International Design Center under grant IDG21300102.}
}


\author{Alberto Costa
}


\institute{A. Costa \at
              Singapore University of Technology and Design \\
              \email{costa@lix.polytechnique.fr}           
}

\date{Received: date / Accepted: date}

\maketitle

\begin{abstract}
A ``quantitative function'' for community detection called modularity density has been proposed by Li, Zhang, Wang, Zhang, and Chen in $[$Phys. Rev. E \textbf{77}, 036109 (2008)$]$. 
{We study the modularity density maximization problem and we discuss some features of the optimal solution. More precisely, we show that in the optimal solution there can be communities having negative modularity density, and we propose a modification of the original formulation to overcome this issue. Moreover, we show that a clique can be divided into two or more parts when maximizing the modularity density. We also compare the solution found by maximizing the modularity density with that obtained by maximizing the modularity on the Zachary karate club network.}
\end{abstract}

\section{Introduction}
\label{intro}

Networks, or graphs, are often used to describe complex systems, and they find application in many fields, e.g., biology and bioinformatics \cite{guimera,PalEtAl05}, recommender systems \cite{Adomavicius05towardthe}, social networks \cite{citeulike:81501}.
One of the topics related to networks which has been studied extensively in the last years is community detection:
given a network $G=(V,E)$, where $V$ is the set of vertices and $E$ is the set of edges, one wants to find subsets of $V$ (called clusters, or communities, or modules) which are more connected with vertices in the same community than with vertices in other communities. Hence, a partition is obtained by splitting $V$ in $m$ communities $\{V_1,\dots,V_m\}$ which cover $V$. In general, these communities are non-empty, non-overlapping, and their number $m$ is not known a priori.

There are many ways to define a community. For example, one may specify some rules that each community must respect \cite{PhysRevE.85.046113,coh,Radicchi}. Another approach is to use some heuristics (see for example \cite{anncosta,citeulike:81501}). Alternatively, one could specify an objective function to maximize or minimize. Concerning the latter, probably the most famous of such functions is modularity, which represents the fraction of edges within communities minus the expected fraction of such edges in a random network with the same degree distribution \cite{citeulike:81501,PhysRevE.69.026113}. More precisely, using the notation of \cite{PhysRevE.77.036109}, modularity is defined as follows:
\begin{equation}
Q=\sum_{i=1}^m\left[\frac{L(V_i,V_i)}{L(V,V)}-\left(\frac{L(V_i,V)}{L(V,V)}\right)^2\right],
\end{equation}
where $L(V_i,V_i)$ is twice the number of edges in the community $V_i$, $L(V,V)$ is twice the number of edges of $G$ (i.e., $2|E|$), and $L(V_i,V)$ is equal to the sum of degrees of vertices belonging to the community $V_i$. Notice that, in order to find a good quality partition, modularity should be maximized.

Although modularity is widely used, it presents some issues, as degeneracy and resolution limit \cite{fortunato-2007-104,PhysRevE.81.046106}. Degeneracy is related to the possible presence of several high modularity partitions which makes it difficult to find the global optimum. Resolution limit refers to the sensitivity of modularity to the total number of edges in the network, hence small communities may not be identified and remain hidden inside larger ones.
To overcome the resolution limit of modularity, a measure called modularity density has been proposed by Li, Zhang, Wang, Zhang, and Chen in \cite{PhysRevE.77.036109}. More precisely, modularity density is defined as:
\begin{equation}
D=\sum_{i=1}^m d(G_i)=\sum_{i=1}^m\left[\frac{L(V_i,V_i)-L(V_i,\bar{V_i})}{|V_i|}\right],
\end{equation}
where $d(G_i)$ is the modularity density associated with the community $V_i$, $L(V_i,\bar{V_i})$ is the number of edges joining a vertex in $V_i$ to a vertex belonging to another community, and $|V_i|$ is the number of vertices belonging to $V_i$. 

{This paper is organized as follows: in Section \ref{sec:discussion} we discuss some properties of the modularity density. In particular, in Section \ref{sec:lb} we show that in the optimal solution there can be communities having a negative modularity density value, and we propose a constraint to overcome this issue. We also show how this constraint can help to derive a mixed integer linear programming reformulation of the problem, and we point out the relationship between this constraint and the weak definition of Radicchi \emph{et. al} \cite{Radicchi}. In Section \ref{sec:split} we show that a clique can be split in the optimal solution. After that, in Section \ref{sec:comm} we comment some wrong and inaccurate statements of \cite{PhysRevE.77.036109}.
Finally, in Section \ref{sec:concl} we present the conclusions.}

\section{Discussion on the properties of modularity density}
\label{sec:discussion}

We discuss in the following some features of modularity density.

\subsection{Lower bound for modularity density of a community}
\label{sec:lb}
\begin{figure}[ht]
\centering
 \includegraphics[scale=0.4]{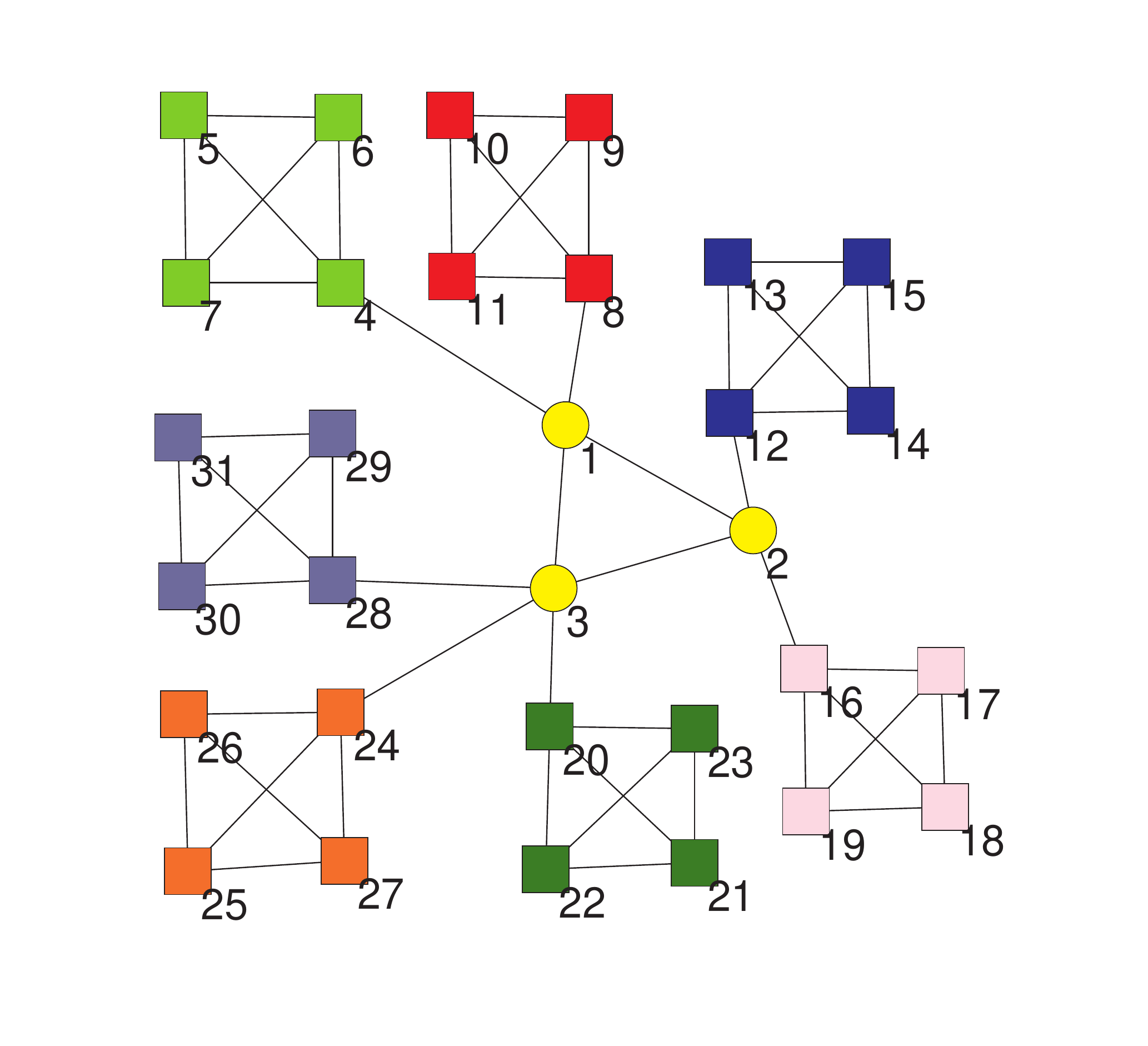}%
\caption{Example of a network for which the optimal solution contains a community with negative modularity density (color online).}
\label{fig:moddneg}
\end{figure}

{As for modularity, one should maximize the modularity density to find a good quality partition. In fact, Li \emph{et. al} \cite{PhysRevE.77.036109} state that ``clearly the maximum $D$ value is often achieved when the network is correctly partitioned''. 
Intuitively, the modularity density of each community should assume a high value, but there are cases where some communities can have negative modularity density value in the optimal solution.} To show this, consider the network with 31 vertices of Fig.~\ref{fig:moddneg}: it consists of 7 cliques, each of them having 4 vertices (square shape), connected to a smaller clique with 3 vertices (circle shape). 
The optimal solution we found by solving the modularity density maximization problem using the exact formulations presented in \cite{modd_ejor} is a partition into 8 communities: 7 communities correspond to the 7 cliques having size 4 (i.e.,  the communities are $V_i=\{4i,4i+1,4i+2,4i+3\},\,\forall i\in\{1,\dots, 7\}$) whereas the other community corresponds to the smaller clique with 3 vertices (i.e., $V_8=\{1,2,3\}$). It could be easily checked that the modularity density $D$ of this partition is 18.9167. More precisely, the modularity density value associated with the community $V_8$ is $-\frac{1}{3}$, and $\frac{11}{4}$ for each of the other 7 communities. 
Hence, we cannot assume that in the optimal solution each community has a non-negative modularity density value. Notice that this property is strictly related to the weak definition suggested by Radicchi \emph {et al.} \cite{Radicchi}. We discuss now this point more in detail. Let $k_i^{in}$ be the number of edges connecting the vertex $v_i$ to other vertices in the same community, and $k_i^{out}$ be the number of edges connecting the vertex $v_i$ to vertices belonging 
to other communities (hence, the degree of the vertex $v_i$ is $k_i=k_i^{in}+k_i^{out}$). According to Radicchi \emph {et al.} \cite{Radicchi}, a subgraph $V_l$ of $V$ is a community in the weak sense if:
\begin{equation}
\label{eq:rweak}
\sum_{v_i\in V_l} k_i^{in} > \sum_{v_i\in V_l} k_i^{out},
\end{equation}
which implies that twice the number of edges inside a community is \emph{strictly} greater than the number of edges connecting a vertex of the community to a vertex in another community (cut edges).
Let $x_{il}$ be a binary variable equal to 1 if the vertex $v_i$ is inside the community $l$, and 0 otherwise, and let $a_{ij}$ be an element of the adjacency matrix of the graph $G$ (i.e., $a_{ij}$ is equal to 1 
if and only if there is an edge connecting $v_i$ to $v_j$). As shown in \cite{coh}, the weak condition \eqref{eq:rweak} is equivalent to:
\begin{equation}
\label{eq:rweak2}
4\sum_{\{v_i,v_j\}\in E} x_{il}x_{jl} \geq \sum_{v_i\in V} k_ix_{il} + 1.
\end{equation}

According to Appendix A of \cite{PhysRevE.77.036109}, modularity density can be expressed as follows:
\begin{equation}
\label{eq:moddformula}
D=\displaystyle\sum_{l=1}^{m}\left(\frac{\displaystyle\sum_{v_i\in V}\sum_{v_j\in V}a_{ij}x_{il}x_{jl}-\displaystyle\sum_{v_i\in V}\sum_{v_j\in V}a_{ij}x_{il}(1-x_{jl})}{\displaystyle\sum_{v_i\in V} x_{il}}\right),
\end{equation}
which can be rewritten as:
\begin{equation}
\label{eq:moddformula2}
D=\displaystyle\sum_{l=1}^{m}\left(\frac{\displaystyle4\sum_{\{v_i,v_j\}\in E}x_{il}x_{jl}-\displaystyle\sum_{v_i\in V}k_ix_{il}}{\displaystyle\sum_{v_i\in V} x_{il}}\right).
\end{equation}
Comparing \eqref{eq:rweak2} and \eqref{eq:moddformula2} it appears that the weak definition is respected if, for each community, the corresponding value of modularity density is \emph{strictly} positive.
Therefore, one could adjoin to the modularity density formulation the constraint \eqref{eq:rweak2} without the +1 on the right-hand side to assure that each community has got a non-negative modularity density value, 
or the constraint \eqref{eq:rweak2} to assure that the partition found is compatible with the weak definition of \cite{Radicchi}. The latter has been studied in \cite{coh} for modularity maximization. 
Let $M=\{1,\dots,m\}$ be the set of the indices of the communities. The binary non-linear formulation which includes the weak constraint can be written as:
\begin{align}
\max \quad& \sum_{l\in M}\left(\frac{ \displaystyle4\sum_{\{v_i,v_j\}\in E}x_{il}x_{jl}-\sum_{v_i\in V}k_ix_{il}}{ \displaystyle\sum_{v_i\in V} x_{il}}\right)\label{2objf3bcq} \\
\text{s.t.}\quad&\forall l\in M\quad 1\leq\sum_{v_i\in V}x_{il}\leq |V|-1\label{eq:2ob2wbcq}\\
&\forall v_i\in V\quad \sum_{l\in M}x_{il}=1\label{appq}\\
&\forall l \in M\quad 4\sum_{\{v_i,v_j\}\in E} x_{il}x_{jl} \geq \sum_{v_i\in V} k_ix_{il} + L\label{weak}\\
      &\forall l\in M,\,\forall v_i\in V\quad x_{il}\in \{0,1\},\label{2bpev82cq}
      \end{align}
where \eqref{eq:2ob2wbcq} ensures that each community is non-empty and that all the vertices are not assigned to the same community (we suppose that there are at least two communities, otherwise the solution would be the trivial partition containing all the vertices), \eqref{appq} imposes
that each vertex belong to only one community, and \eqref{weak} is the weak constraint, where the value of $L$ is 1 if we consider the original definition in \cite{Radicchi} and 0 if we only want to guarantee
that each community assumes a non-negative value of modularity density.

The advantage of this new formulation, which will be discussed in the following, is that we can derive a more efficient exact linearization of the objective function. As noticed in \cite{modd_ejor}, the difficult part is the linearization of the fractions
arising in \eqref{2objf3bcq} (the products $x_{il}x_{jl}$ involving two binary variables can be easily linearized exactly using the Fortet inequalities \cite{fortet} or the dual approach presented in \cite{costasea}).
To ease the explanation, we consider the modularity density of the community $V_l$ (the same technique can be applied to linearize the modularity density of all the other communities).
The idea used in \cite{modd_ejor} for the linearization the modularity density of $V_l$ (formulation MDL) is the following.
First, we introduce a variable $\alpha_l$ representing the modularity density of $V_l$:
\begin{equation}
\label{eq:ref1}
\alpha_l=\frac{\displaystyle4\sum_{\{v_i,v_j\}\in E}x_{il}x_{jl}-\displaystyle\sum_{v_i\in V}k_ix_{il}}{\displaystyle\sum_{v_i\in V} x_{il}}.
\end{equation}
Thanks to the fact that empty communities are not allowed (see constraint \eqref{eq:2ob2wbcq}), the denominator of \eqref{eq:ref1} is greater than 0, hence we can write: 
\begin{equation}
\label{eq:ref2}
\displaystyle4\sum_{\{v_i,v_j\}\in E}x_{il}x_{jl}-\displaystyle\sum_{v_i\in V}k_ix_{il}=\displaystyle\sum_{v_i\in V} \alpha_lx_{il}.
\end{equation}
We need now to linearize each product $\alpha_lx_{il}$. We can derive an exact linearization by means of the McCormick inequalities \cite{mccormick}, because $x_{il}$ is binary. However,
we need a lower and an upper bound on $\alpha_l$. Indeed, the tighter those bounds, the better the linearization. Concerning the upper bound, it has been computed in \cite{modd_ejor} by solving
an auxiliary problem, whereas a theoretical lower bound $L_\alpha=-\frac{k_{\max_1}+k_{\max_2}}{2}$ has been employed (where $k_{\max_1}$ and $k_{\max_2}$ are two vertices with the highest degrees).
If constraint \eqref{weak} holds then the lower bound for $\alpha_l$ would be $L=1$ or $L=0$ (depending on the value of $L$ in \eqref{weak}).
Those values provide in general a lower bound which is 
much tighter than $L_\alpha=-\frac{k_{\max_1}+k_{\max_2}}{2}$, and which does not depend on the size of the instances (on the other hand, the quality of the bound $L_\alpha=-\frac{k_{\max_1}+k_{\max_2}}{2}$ decreases
with the size of the instance, in general). This idea can be also extended to the formulation MDB in \cite{modd_ejor}, where a binary decomposition of the denominator of \eqref{eq:ref1} has been
employed to decrease the number of products to linearize with the McCormick inequalities.

Using the formulation \eqref{2objf3bcq}-\eqref{2bpev82cq} with both $L=0$ and $L=1$, the partition into 8 communities represented in Fig.~\ref{fig:moddneg} is infeasible. The optimal solution is found when
the number of communities is 7: the difference with respect to the previous case is that the clique $\{1,2,3\}$ is in the same community of one of the cliques having size 4 connected to vertex 3 (which one does not matter, the solution would be symmetric).
The modularity density value associated with this new partition is 18.5.

\subsection{Splitting of a clique in the optimal solution}
\label{sec:split}
\begin{figure}[ht]
\centering
 \includegraphics[scale=0.4]{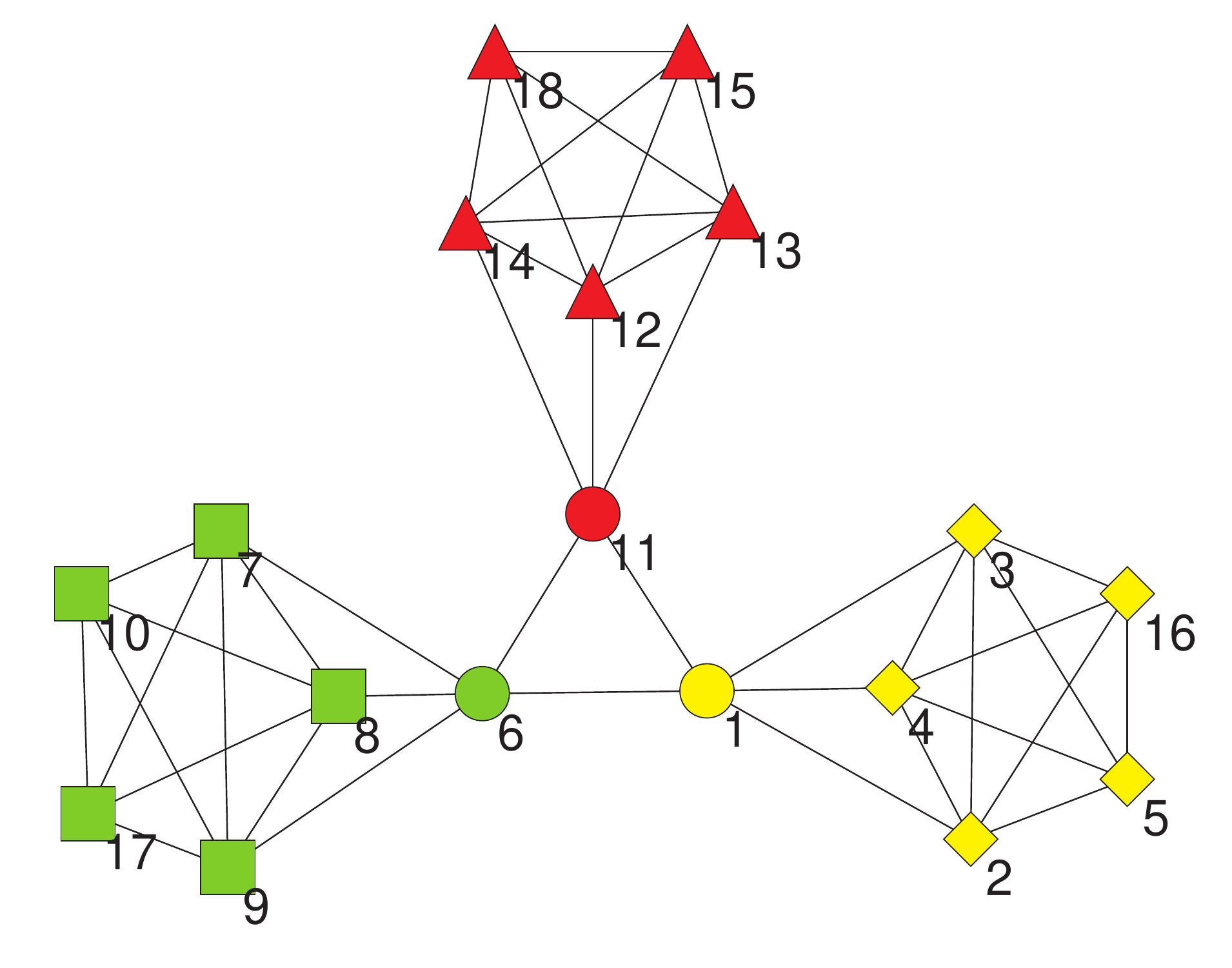}%
\caption{Example of a network composed by three cliques with 5 vertices each, connected to a smaller clique having 3 vertices (color online).}
\label{fig:pr3}
\end{figure}

Consider the network with 18 vertices presented in Fig.~\ref{fig:pr3}.

Maximizing the modularity density when the number of communities is 4 results in a partition consisting of the 4 cliques:
$V_1=\{1,6,11\}$ (circle shape), $V_2=\{2,3,4,5,16\}$ (diamond shape), $V_3=\{7,8,9,10,17\}$ (square shape), $V_4=\{12,13,14,15,18\}$ (triangle shape), and the corresponding modularity density value $D$ is equal to 9.2.
However, a higher value of modularity density is obtained when the number of communities is 3. The partition found is the following: $V_1=\{1,2,3,4,5,16\}$ (yellow color), $V_2=\{6,7,8,9,10,17\}$ (green color), $V_3=\{11,12,13,14,15,18\}$ (red color), and the corresponding value of modularity density is $D=12$. Hence, in the optimal solution the small clique $\{1,6,11\}$ is split among the three other cliques.
Notice that the solution with 4 communities would have been infeasible if using the formulation with the weak constraint \eqref{2objf3bcq}-\eqref{2bpev82cq}, regardless of the value of $L$.

\section{Comment on ``Quantitative function for community detection''}
\label{sec:comm}
Among the properties presented by Li \emph{et al.} \cite{PhysRevE.77.036109}, some of them are not proved, wrong, or need to be clarified.
Discussing and commenting these properties is the subject of this section.

%
%
%
%
%
%
{
\subsection{Non-negative modularity density}
\label{sec:nonneg}
Li \emph{et al.} claim that ``Since our purpose is to maximize the modularity density $D$, every term $d(G_i)$ must be non-negative''. Indeed, this is intuitive, as one may expect that the maximum value of $D$ is obtained when all the terms $d(G_i)$ assume high values. Nevertheless, this is not always true when the number of communities is non-optimal (where the optimal number of communities is that of the partition yielding the highest value of modularity density).
Consider for example the journal index network tested in Section V.~3 of \cite{PhysRevE.77.036109}. The optimal number of communities is 4. However, when trying to maximize the modularity density with 5 communities, the authors state that ``When we intend to split the network into five modules, we get essentially the same partition as with four, only with the singly connected journal Conservation Biology split off by itself as a community''. It is easy to check that the modularity density value of the community consisting only of the vertex associated to the journal Conservation Biology is -1.
Actually, even when the number of communities is optimal, the property could not hold: in some cases having a community with a small negative value of modularity density allows other communities to assume higher modularity density values, thus yielding a higher value of $D$, as shown in Section \ref{sec:lb}.
Notice that this wrong statement can yield wrong formulations for the modularity density maximization problem. As pointed out in Section \ref{sec:lb}, in \cite{modd_ejor} some exact linearizations of the non-linear formulation proposed in \cite{PhysRevE.77.036109} are introduced, and they require a lower bound on the modularity density value. Using 0, as suggested in \cite{PhysRevE.77.036109}, would produce a wrong model. Moreover, the statement ``the partition (subgraphs) by optimizing $D$ results in communities consistent with the weak definition suggested by Radicchi \emph {et al.}'' is also not correct.
To summarize, there are two mistakes in their statements:
\begin{itemize}
 \item it is not true that modularity density for a community is always non-negative in the optimal solution (as shown in Fig.~\ref{fig:moddneg});
 \item even though modularity density was non-negative for all communities in the optimal solution, this would not be enough to assure that the weak condition holds, because that condition
       requires the modularity density to be \emph{strictly} positive for all communities (this because of the strict inequality in \eqref{eq:rweak}, that yields the +1 on the right-hand side of \eqref{eq:rweak2}).
\end{itemize}
}

{
\subsection{Division of cliques in the optimal solution}
\label{sec:commsplit}
One of the properties presented by Li \emph{et al.} is  ``Given a clique with $n$ vertices, maximizing modularity density or $D$ does not divide it into two or more parts''. This statement should be clarified: the proof of the authors assumes the whole network being a clique (i.e., the clique has no external edges connecting it with other vertices), and it does not refer to any clique which can be found in a network (even though this property is later employed to prove some other results for networks containing some cliques, see Fig.~1 and Sections III.~B-C in \cite{PhysRevE.77.036109}). In fact, if the clique is densely connected to external vertices, it could be split, as shown in Section \ref{sec:split}
}

\subsection{Complexity of modularity density maximization}
\label{sec:compl}
Li \emph{et al.} state that ``The search for optimal modularity density $D$ is a \textbf{NP}-hard problem due to the fact that the space of possible partitions grows faster than any power of system size''. This is not an appropriate definition of \textbf{NP}-hardness. Consider for example the shortest path problem \cite{Cormen:2009:IAT:1614191}: even though the space of the possible solutions is exponential, the problem belongs to \textbf{P}. This does not mean that modularity density maximization is not a \textbf{NP}-hard problem, but the correctness of this statement should be proven in a more appropriate way, for example by means of a reduction from a \textbf{NP}-complete problem to the decision version of the modularity density maximization (as done for modularity \cite{citeulike:4363443}).
Notice that some papers already cite \cite{PhysRevE.77.036109} as reference for the \textbf{NP}-hardness of modularity density maximization \cite{6785984,OSB}.

\subsection{Wrong result for Zachary karate club network}
\label{sec:zac}
\begin{figure}[ht]
\centering
 \includegraphics[scale=0.4]{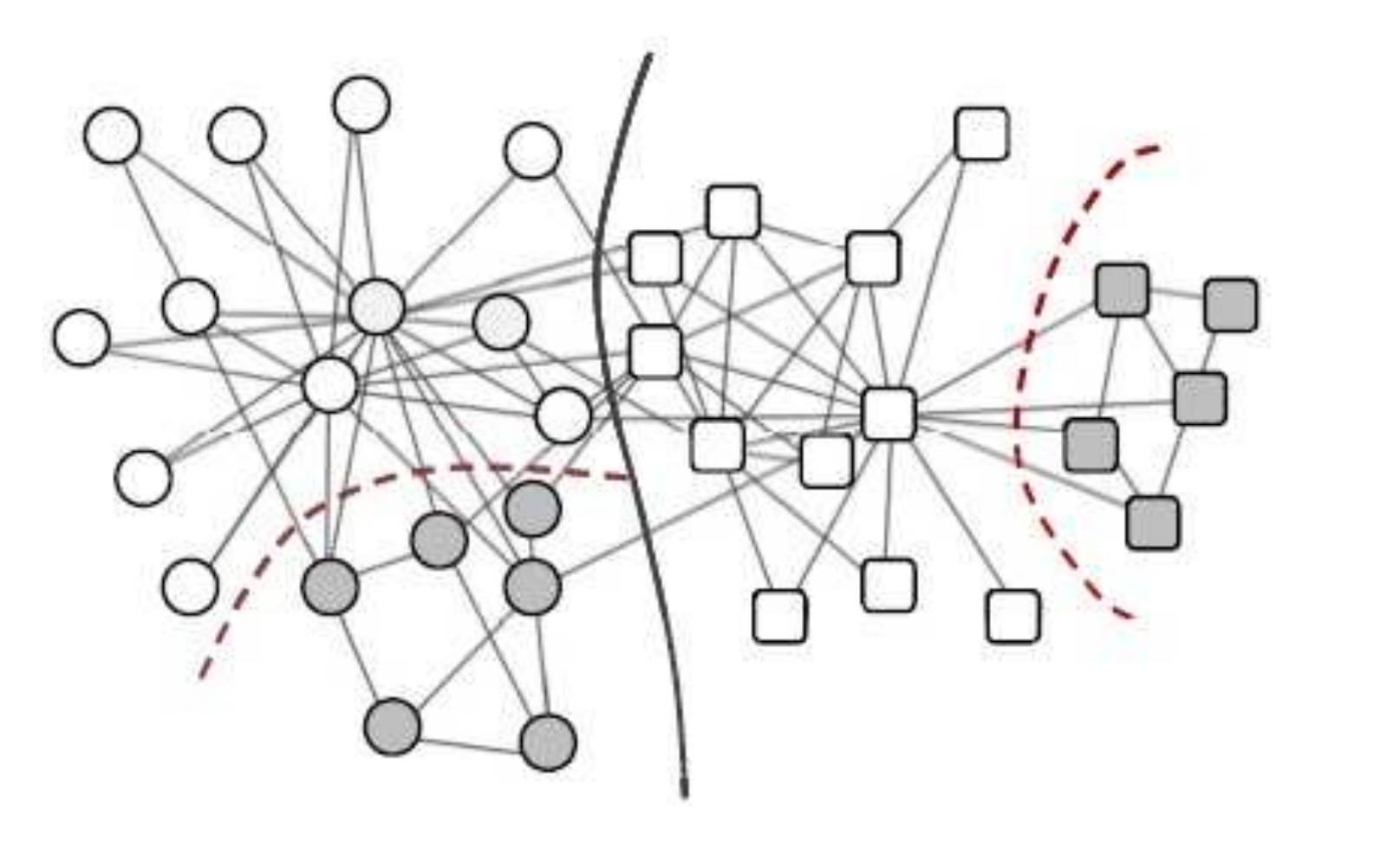}%
\caption{Partitions into 2 and 4 communities of the Zachary karate club network presented in \cite{PhysRevE.77.036109} (color online).}
\label{fig:zac2}
\end{figure}

\begin{figure}[ht]
\centering
 \includegraphics[scale=0.4]{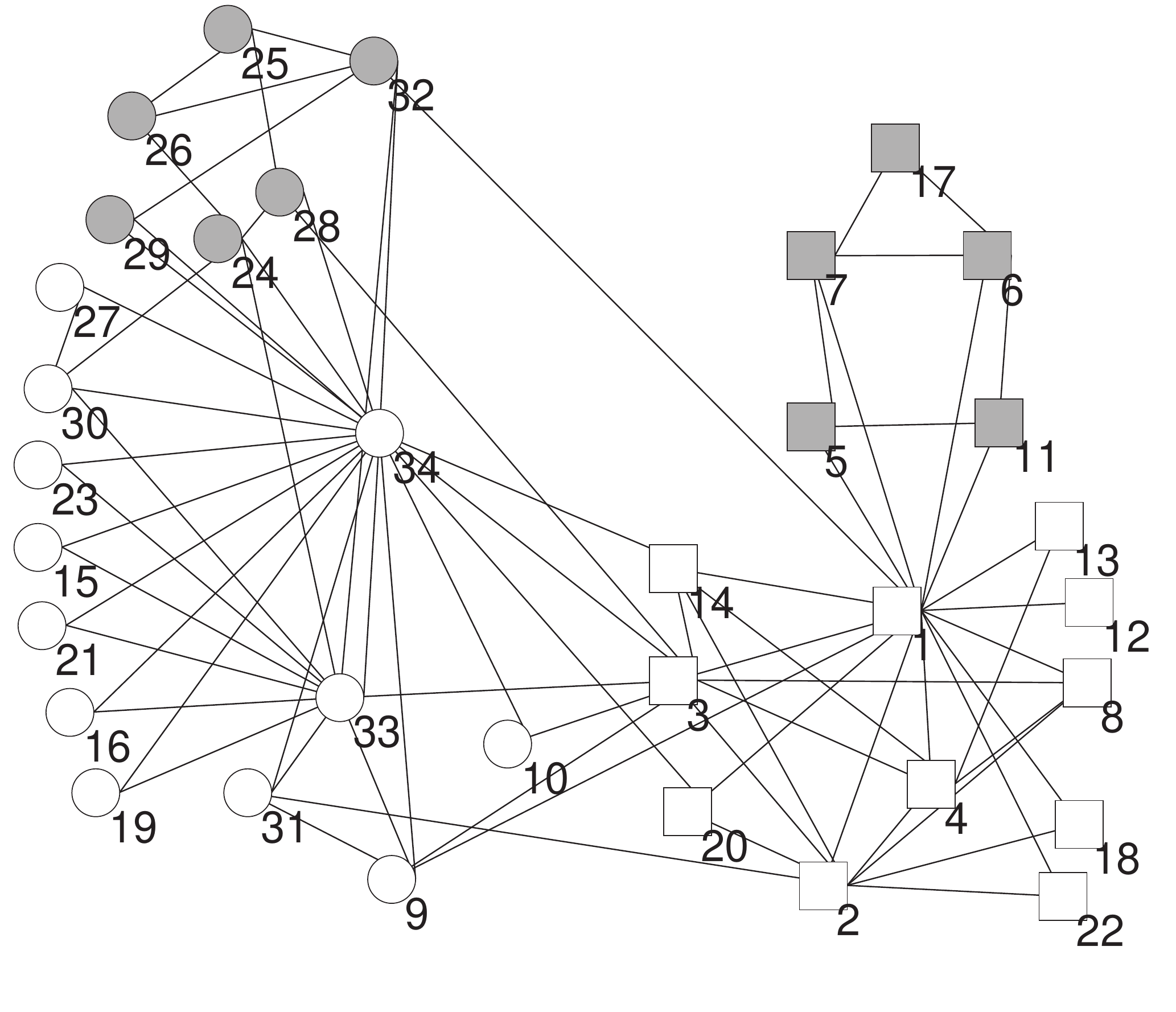}%
\caption{Same partitions represented in Fig.~\ref{fig:zac2} with labels for the vertices (color online).}
\label{fig:zac}
\end{figure}

In Section V Li \emph{et al.} test their function with some artificial and real-world instances. Concerning the latter, they present the results for the famous Zachary karate club network \cite{zachary}.
Commenting on the solution found, the authors claim that ``By using our method, the network was partitioned into two communities exactly consistent with real partition when $k=2$ (see Fig.~3). However, maximizing the $D$ value, we obtained the ``optimal'' partition with $k=4$ which is also reasonable from the topology of the network''.
We now discuss more in detail this point. We show in Fig.~\ref{fig:zac2} (that is Fig.~3 borrowed from \cite{PhysRevE.77.036109}) the partitions into 2 and 4 communities presented by the authors, and in Fig.~\ref{fig:zac} the same partitions with, in addition, the indications of the labels for the vertices.

We tried to optimize the $D$ value on the Zachary karate club network, but we obtained different results.
The partition with 2 communities found by Li \emph{et al.} consists of $V_1=\{1,2,3,4,5,6,7,8,11,12,13,14,17,18,20,22\}$ (squares) and $V_2=\{9,10,15,16,19,21,23,24,25,26,27,28,29,30,31,32,33,34\}$ (circles).
When the number of communities is 4, each community of the previous partition is further split in two. The result is a partition composed by $V_1=\{5,6,7,11,17\}$ (dark squares), $V_2=\{1,2,3,4,8,12,13,14,18,20,22\}$ (white squares), $V_3=\{24,25,26,28,29,32\}$ (dark circles), and $V_4=\{9,10,15,16,19,\\21,23,27,30,31,33,34\}$ (white circles).
We solved the problem of modularity density maximization for the Zachary karate club network with 2, 3, and 4 communities using the exact formulation presented in \cite{modd_ejor}.
The result obtained with 2 communities is consistent with that of the authors, and the value of $D$ is 6.83333.
The result obtained with 3 communities is the following partition: $V_1=\{1,2,3,4,8,10,12,13,14,18,20,22\}$, $V_2=\{9,15,16,19,21,23,24,25,26,\\27,28,29,30,31,32,33,34\}$, and $V_3=\{5,6,7,11,17\}$, with $D=7.8451$. Notice that this is the same partition obtained with the almost-strong rule \cite{PhysRevE.85.046113}. Finally, the partition into 4 communities gave a different result from that of the authors. The solution we found is the following: 
$V_1=\{5,6,7,11,17\}$, $V_2=\{1,2,3,4,8,12,13,14,18,20,22\}$, $V_3=\{25,26,29,32\}$, and $V_4=\{9,10,15,16,\\19,21,23,24,27,28,30,31,33,34\}$, with a value of $D$ of 7.54481. The difference with respect to the solution of the authors is that vertices 24 and 28 are moved from the community $V_3$ to $V_4$. 
Actually, if vertices 24 and 28 belong to $V_3$, the corresponding value of $D$ is 7.50909, that is non-optimal when there are 4 communities. These results are summarized in Table~\ref{tab:res}, together with the values of modularity $Q$ associated with the partition found by maximizing the modularity density. 
\begin{table}[h!]
\centering
\caption{Results obtained by maximizing the modularity density $D$ on the Zachary karate club network, and corresponding values of modularity $Q$. The results refer to the optimal partitions obtained with 2, 3, and 4 communities, as well as to the non-optimal partition into 4 communities presented by the authors in \cite{PhysRevE.77.036109} (see Fig.~\ref{fig:zac2}-\ref{fig:zac}).}
\label{tab:res}
\begin{tabular}{ccc}
\hline
m&$D$&$Q$\\\hline
2 & 6.83333 & 0.371466\\\hline
3 & \textbf{7.8451} & 0.402038\\\hline
4 & 7.54481 & 0.415105\\\hline
4 (Fig.~\ref{fig:zac2}-\ref{fig:zac}) & 7.50909 & \textbf{0.41979}\\\hline
\end{tabular}
\end{table}
Looking at Table~\ref{tab:res}, we can also notice that the solution with 4 communities of Fig.~\ref{fig:zac2} corresponds to the highest value of modularity $Q$, whereas the best solution found by maximizing $D$ with 4 communities is different, as explained earlier. To summarize, not only the solution proposed in \cite{PhysRevE.77.036109} is non-optimal with respect to the number of communities (which should be 3 instead of 4), but even fixing the number of communities at 4 their solutions is non-optimal with respect to the modularity density value (which should be 0.754481 instead of 0.750909).
The reason for this behavior is that all the results presented in \cite{PhysRevE.77.036109} are based on a method which finds only local optima (as confirmed by the authors), hence there is no guarantee of global optimality.

\section{Conclusion}
\label{sec:concl}
In this paper {we have discussed some properties of modularity density, and we have shown the relationship with the weak definition of community of Radicchi \emph{et. al} \cite{Radicchi}. This remark allowed us to derive a new formulation, which is easier to linearize and which ensures that in the optimal solution each community has a non-negative value of modularity density. 
Moreover, we have clarified, commented, and corrected some wrong and inaccurate statements presented in \cite{PhysRevE.77.036109}. Despite these issues, modularity density remains a very interesting criterion, due to its capability of fixing the resolution limit issue of modularity. For this reason, we targeted our effort to a better characterization and description of its features.}

\bibliographystyle{spmpsci}      
\bibliography{bibliografia}

%
%

\end{document}